\documentclass[journal]{IEEEtran}

\usepackage[usenames]{color}
\usepackage{balance}
\usepackage{amsmath}
\usepackage{amssymb}
\usepackage{multirow}
\usepackage{cite}
\usepackage{ifpdf}
\usepackage{hyperref}
\usepackage{fancyhdr}

\ifCLASSINFOpdf
   \usepackage[pdftex]{graphicx}
   \graphicspath{{./pdf/}}
   \DeclareGraphicsExtensions{{.pdf}}
\else
   \usepackage[dvips]{graphicx}
   \graphicspath{{./eps/}}
   \DeclareGraphicsExtensions{{.eps}}
\fi
\begin{document}

\title{An Analysis of the DS-CDMA Cellular Uplink \\for Arbitrary and Constrained Topologies}
\author{Don Torrieri,~\IEEEmembership{Senior~Member,~IEEE,} Matthew C.
Valenti,~\IEEEmembership{Senior~Member,~IEEE,} \\ and Salvatore
Talarico,~\IEEEmembership{Student Member,~IEEE.} \thanks{Portions of this work
were presented at the IEEE International Conference on
Communications (ICC), Budapest, Hungary, June 2013.} \thanks{M.~C.~Valenti's
work was sponsored by the National Science Foundation under Award No.
CNS-0750821 and by the United States Army Research Laboratory under Contract
W911NF-10-0109.} \thanks{D. ~Torrieri is with the US Army Research Laboratory,
Adelphi, MD (email: don.j.torrieri.civ@mail.mil).} \thanks{M.~C.~Valenti and
S.~Talarico are with West Virginia University, Morgantown, WV, U.S.A. (email:
valenti@ieee.org, Salvatore.Talarico81@gmail.com).}
\thanks{Digital Object Identifier 10.1109/TCOMM.2013.13.120911} }
\date{}


\pagestyle{fancy}
\fancyhead[RO,LE]{\small\thepage}
\fancyhead[LO]{\small IEEE TRANSACTIONS ON COMMUNICATIONS, ACCEPTED FOR PUBLICATION}
\fancyhead[RE]{\small TORRIERI et al.: AN ANALYSIS OF THE DS-CDMA CELLULAR UPLINK FOR ARBITRARY AND CONSTRAINED TOPOLOGIES}
\fancyfoot[L,R,C]{}
\renewcommand{\headrulewidth}{0pt}

\maketitle

\begin{abstract}
A new analysis is presented for the direct-sequence code-division multiple
access (DS-CDMA) cellular uplink. For a given network topology, closed-form
expressions are found for the outage probability and rate of each uplink in
the presence of path-dependent Nakagami fading and shadowing. The topology may
be arbitrary or modeled by a random spatial distribution with a fixed number of
base stations and mobiles placed over a finite area. The analysis is more
detailed and accurate than existing ones and facilitates the resolution of
network design issues including the influence of the minimum base-station
separation, the role of the spreading factor, and the impact of various
power-control and rate-control policies. It is shown that once power control
is established, the rate can be allocated according to a fixed-rate or
variable-rate policy with the objective of either meeting an outage constraint
or maximizing throughput. An advantage of variable-rate power control is that
it allows an outage constraint to be enforced on every uplink, which is
impossible when a fixed rate is used throughout the network.

\end{abstract}

\begin{IEEEkeywords}
CDMA, cellular network, uplink, power control, rate control
\end{IEEEkeywords}

\thispagestyle{empty}

\section{Introduction}

\PARstart{T}{he} classical analysis of the cellular uplinks (e.g., \cite{gil,vit,zor})
in a cellular network entails a number of
questionable assumptions, including the existence of a
lattice or regular grid of base stations and the modeling of intercell
interference at a base station as a fixed fraction of the total interference.
Although conceptually simple and locally tractable, the grid assumption is a
poor model for actual base-station deployments, which cannot assume a regular
grid structure due to a variety of regulatory and physical constraints. The
intercell-interference assumption is untenable because the fractional
proportion of intercell interference varies substantially with the mobile and
base-station locations, the shadowing, and the fading.

More recent analyses of cellular networks (e.g.,
\cite{and,Dhillon:2012,Blas:2013}) locate the mobiles and/or the base stations
according to a two-dimensional Poisson point process (PPP) over a network that
extends infinitely on the Euclidian plane, thereby allowing the use of
analytical tools from stochastic geometry \cite{stoy,bacc,web}.
Although the two principal limitations of the classical approach
are eliminated, the PPP approach is still unrealistic because it does not
permit a minimum separation between base stations or between mobiles, although
both minimum separations are characteristic of actual macro-cellular
deployments. Other problems with the PPP model are that no network has an
infinite area and that any realization of a PPP could have an implausibly
large number of mobiles or base stations placed within a finite area.

The cellular downlink is considered in \cite{and} and \cite{Blas:2013}, while
the cellular uplink is considered in \cite{Dhillon:2012}. The model for the
uplink proposed in \cite{Dhillon:2012} involves drawing the locations of the
mobiles from a PPP and then locating the base station for each mobile
uniformly in the mobile's Voronoi cell. This model supposes a one-to-one
relationship between mobiles and base stations and therefore cannot be used to
model networks that allow several simultaneous connections per base station.
In \cite{Blas:2013}, it is shown that when the variance of the lognormal
shadowing is large, the PPP model is a reasonable approximation even for a
network with base stations located on a regular lattice. However, in these
works the shadowing and fading are lumped into a single variable, which
confounds their relative contributions. In particular, shadowing affects the
base-station selection, but fading does not. A key motivation of the present
paper is to separately account for the shadowing and fading.

The goal of this paper is to provide a new and accurate analysis of the
direct-sequence code-division multiple-access (DS-CDMA) uplink. The analysis
can be used to determine the exact outage probability of an arbitrary fixed
topology with no need for simulation. In addition, a constrained spatial model
is introduced, which imposes a minimum separation among base stations. The
minimum separation can be selected to match the observed locations of an
actual cellular deployment. Analytical techniques are presented for
characterizing the performance of networks drawn from the constrained spatial
model. The model facilitates the observation of certain phenomena. For
instance, it is shown subsequently in Fig. \ref{Figure:TCRbs} that the minimum
separation between base stations has a significant effect on area spectral
efficiency. Another benefit of the analysis is that it exposes the tradeoffs
entailed in choosing among various power-control and rate-control policies in
terms of both the outage probabilities and the area spectral efficiencies.

The analysis in this paper applies a closed-form expression \cite{tor2} for
the\ \emph{conditional} outage probability of a communication link, where the
conditioning is with respect to an arbitrary realization of the network
topology. The techniques of \cite{tor2} are adapted to provide the uplink
outage probability of an arbitrary DS-CDMA uplink in closed form with no need
to simulate the corresponding channels. A Nakagami-m fading model is assumed,
which models a wide class of channels, and the fading parameters do not need
to be identical for all communication links. This flexibility allows the
modeling of distance-dependent fading, where mobiles close to the base station
have a dominant line-of-sight path, but the more distant mobiles do not.

Although our analysis is applicable for arbitrary topologies, we focus on
modeling the network with a constrained spatial model. The spatial model
places a fixed number of base stations within a region of finite extent. The
model enforces a minimum separation among the base stations for each
\emph{network realization}, which comprises a base-station placement, a mobile
placement, and a shadowing realization. The model for both mobile and
base-station placement is the \emph{uniform-clustering} model \cite{tor2},
which entails a uniformly distributed placement within the network after
certain regions around mobiles and base stations have been excluded. To characterize the performance of the
constrained spatial model, realizations of the network are drawn according to
the desired spatial and shadowing models, and the computed outage
probabilities are collected. These outage probabilities can be used to
characterize the average uplink performance.
Alternatively, the outage probability of each
uplink can be constrained, and the statistics of the rate provided to each
mobile for its uplink can be determined under various
resource-allocation policies.

A DS-CDMA uplink differs from a downlink, which has been presented in a
separate paper \cite{val}, in at least four significant ways. First, the
sources of interference are many mobiles for the uplink whereas the sources
are a few base stations for the downlink. Second, the uplink signals arriving
at a base station are asynchronous. Therefore, they are not orthogonal and
intracell interference can not be ignored. Third, sectorization is a critical
factor in uplink performance whereas it is
of minor
importance in downlink performance. Fourth, because base stations are equipped
with better transmit high-power amplifiers and receive low-noise amplifiers
than the mobiles, the operational signal-to-noise ratio is typically 5-10 dB
lower for the uplink than for the downlink.

The remainder of this paper is organized as follows. Section
\ref{Section:SystemModel} presents a model of the network culminating in an
expression for SINR in Section \ref{Section:SINR}. Section \ref{Section:Outage} adapts the expression for
outage probability published in \cite{tor2} to the analysis of an arbitrary
DS-CDMA uplink. Section \ref{Section:Policies} discusses policies for power
control, rate control, and cell association/reselection. A performance
analysis given in Section \ref{Section:Performance} compares several network
policies on the basis of outage probability, throughput, area spectral
efficiency, and fairness. The section also
investigates the influence of the spreading factor, minimum base-station
separation, and base-station reselection policy.

\section{Network Model}

\label{Section:SystemModel}

A \textit{sector} is defined as the range of
angles from which a directional sector antenna can receive signals. A mobile
within the sector of a sector antenna is said to be \textit{covered} by the
sector antenna. Cells may be divided into sectors by using several directional
sector antennas or arrays, each covering disjoint angles, at the base
stations. Only mobiles in the directions covered by a sector antenna can cause
intracell or intercell multiple-access interference on the \textit{uplink}
from a mobile to its associated sector antenna. Thus, the number of
interfering signals on an uplink is reduced approximately by a factor equal to
the number of sectors. Practical sector antennas have patterns with sidelobes
that extend into adjacent sectors, but the performance degradation due to
overlapping sectors is significant only for a small percentage of mobile
locations. Three ideal sector antennas and sectors per base station, each
covering $2\pi/3$ radians, are assumed in the subsequent analysis.\ The mobile
antennas are assumed to be omnidirectional.

The network comprises $C$ base stations and cells, $3C$ sectors $\{S_{1}%
,...,S_{3C}\},$ and $M$ mobiles $\{X_{1},...,X_{M}\}$. The base stations and
mobiles are confined to a finite area, which is assumed to be a circle of
radius $r_{\mathsf{net}}$ and area $\pi r_{\mathsf{net}}^{2}$. The sector
boundary angles are the same for all base stations. The variable $S_{j}$
represents both the $j^{th}$ sector antenna and its location, and $X_{i}$
represents the $i^{th}$ mobile and its location.

An \emph{exclusion zone} of radius $r_{\mathsf{bs}}$ surrounds each base
station, and no other base stations are allowed within this zone. Similarly,
an exclusion zone of radius $r_{\mathsf{m}}$ surrounds each mobile where no
other mobiles are allowed. The minimum separation between mobiles is generally
much smaller than the minimum separation between the base stations.
Base-station exclusion zones are primarily determined by economic
considerations and the terrain, whereas mobile exclusion zones are determined
by the need to avoid physical collision. The value of $r_{bs}$
could be empirically determined by examining the locations of base stations in a real
network. The value of $r_{bs}$ could be selected to provide the best statistical fit.
This procedure is similar to determining values
for fading and shadowing parameters by statistically fitting actual data.

\begin{figure}[ptb]%

\centering
\vspace{0.15cm}
\includegraphics[width=8cm]{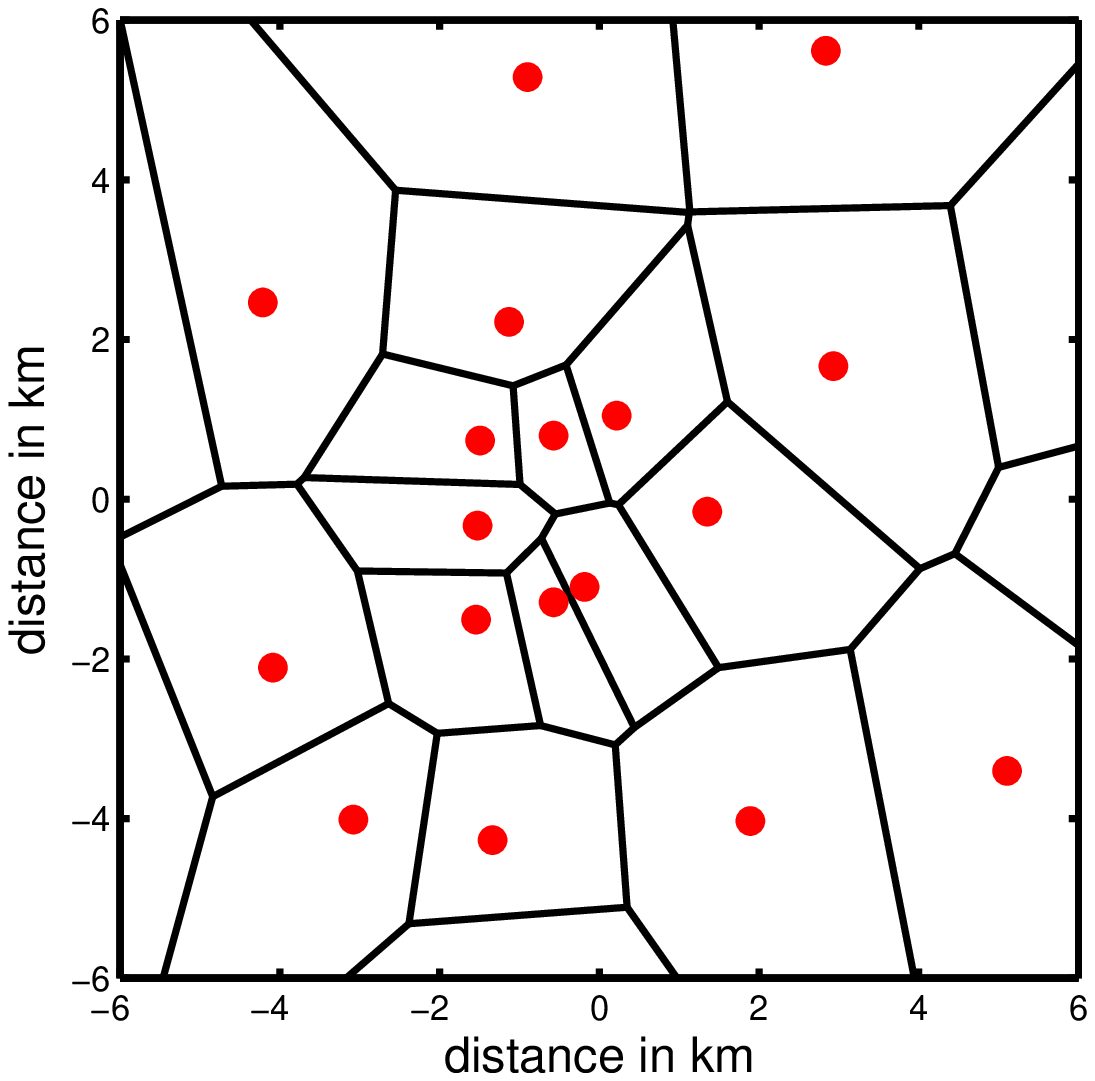}
\vspace{0.25cm}

\hspace{-0.1cm}\includegraphics[width=7.5cm]{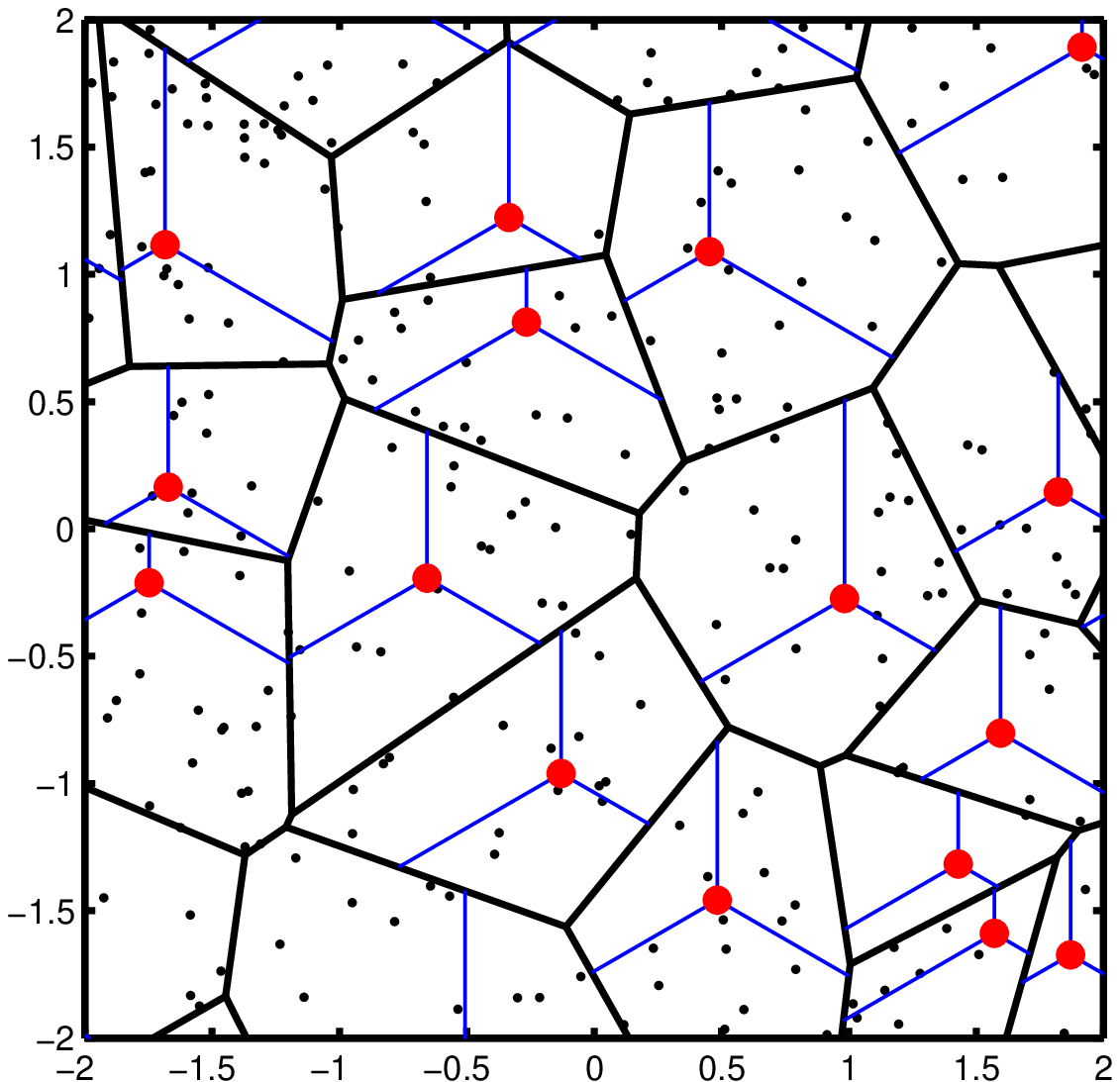}
\caption{ Example network topologies. Base stations are represented by large
circles, and cell boundaries are represented by thick lines. Top
subfigure: Actual base station locations from a current cellular deployment.
Bottom subfigure: Simulated base-station locations using a base-station
exclusion zone $r_{\mathsf{bs}}=0.25$. In the bottom subfigure, the simulated
positions of the mobiles are represented by small dots, the sector boundaries
are represented by light lines, and the average cell load is $M/C=16$
mobiles.
} \label{Figure:Network}

\vspace{-0.75cm}

\end{figure}

Example network topologies are shown in Fig. \ref{Figure:Network}. The top
subfigure shows the locations of actual base stations in a small city with
hilly terrain. The base-station locations are given by the large filled
circles, and the Voronoi cells are indicated in the figure. The minimum
base-station separation is observed to be about 0.43 km. The bottom subfigure
shows a portion of a randomly generated network with average number
of mobiles per cell $M/C=16$, a base-station exclusion radius $r_{\mathsf{bs}%
}=0.25$, and a mobile exclusion radius $r_{\mathsf{m}}=0.01$. The locations of
the mobiles are represented by small dots, and lines indicate the
angular coverage of sector antennas.
The topological similarity of the two subfigures supports the use of the uniform-clustering model.

\section{SINR}

\label{Section:SINR}

Consider a reference receiver of a sector antenna that receives a desired
signal from a reference mobile within its cell and sector. Both intracell and
intercell interference are received from other mobiles within the covered
angle of the sector, but interference from mobiles in extraneous sectors is
negligible.
The varying propagation delays from the interfering mobiles cause
their interference signals to be asynchronous with respect to the desired signal.

In a DS-CDMA network of asynchronous quadriphase direct-sequence systems, a
multiple-access interference signal with power $I$ before despreading is
reduced after despreading to the power level $Ih(\tau_{o})/G$, where $G$ is the
processing gain or spreading factor, and $h(\tau_{o})$ is a function of the
chip waveform and the timing offset $\tau_{o}$ of the interference spreading
sequence relative to that of the desired or reference signal. If $\tau_{o}$ is
assumed to have a uniform distribution over [0, $T_{c}],$ then the expected
value of $h(\tau_{o})$ is the chip factor $h$. For rectangular chip
waveforms, $h=2/3$ \cite{tor}, \cite{tor3}. It is assumed henceforth that
$G/h(\tau_{o})$ is a constant equal to $G/h$ at each sector receiver.

Let $\mathcal{A}_{j}$ denote the set of mobiles \emph{covered} by sector
antenna $S_{j}$. A mobile $X_{i}\in\mathcal{A}_{j}$ will be \emph{associated}
with $S_{j}$ if the mobile's signal is received at $S_{j}$ with a higher
average power than at any other sector antenna in the network. Let
${\mathcal{X}}_{j}\subset\mathcal{A}_{j}$ denote the set of mobiles associated
with sector antenna $S_{j}$. Let $X_{r}\in\mathcal{X}_{j}$ denote a reference
mobile that transmits a desired signal to $S_{j}$. The power of $X_{r}$
received at $S_{j}$ is not significantly affected by the spreading factor and
depends on the fading and path-loss models. The power of $X_{i},$ $i\neq r,$
received at $S_{j}$ is nonzero only if $X_{i}\in\mathcal{A}_{j},$ is reduced
by $G/h,$ and also depends on the fading and path-loss models. We assume that
path loss has a power-law dependence on distance and is perturbed by
shadowing. When accounting for fading and path loss, the despread
instantaneous power of $X_{i}$ received at $S_{j}$ is
\begin{equation}
\rho_{i,j}=%
\begin{cases}
{P}_{r}g_{r,j}10^{\xi_{r,j}/10}f\left(  ||S_{j}-X_{r}||\right)  & i=r\\
\left(  \frac{h}{G}\right)  {P}_{i}g_{i,j}10^{\xi_{i,j}/10}f\left(
||S_{j}-X_{i}||\right)  & \hspace{0cm}i:X_{i}\in\mathcal{A}_{j}\backslash
X_{r}\\
0 & i:X_{i}\notin\mathcal{A}_{j}%
\end{cases}
\label{eqn:power}%
\end{equation}
where $g_{i,j}$ is the power gain due to fading, $\xi_{i,j}$ is a
\textit{shadowing factor}, ${P}_{i}$ is the power transmitted by $X_{i}$, and
$\mathcal{A}_{j}\backslash X_{r}$ is set $\mathcal{A}_{j}$ with element
$X_{r}$ removed.
The
\{$g_{i,j}\}$ are independent with unit-mean, and $g_{i,j}=a_{i,j}^{2}$, where
$a_{i,j}$ has a Nakagami distribution with parameter $m_{i,j}$. While the
$\{g_{i,j}\}$ are independent from each mobile to each base station, they are
not necessarily identically distributed, and each link can have a distinct
$m_{i,j}$. When the channel between $S_{j}$ and
$X_{i}$ experiences Rayleigh fading, $m_{i,j}=1$ and $g_{i,j}$ is
exponentially distributed. In the presence of lognormal shadowing, the
$\{\xi_{i,j}\}$ are i.i.d. zero-mean Gaussian random variables with variance
$\sigma_{s}^{2}$. In the absence of shadowing, $\xi_{i,j}=0$. \ The path-loss
function is expressed as the attenuation power law
\begin{equation}
f\left(  d\right)  =\left(  \frac{d}{d_{0}}\right)  ^{-\alpha}
\hspace{-0.4cm},
\hspace{0.5cm}
\text{ \ }d\geq
d_{0},  \label{eqn:pathloss}%
\end{equation}
where $\alpha\geq2$ is the attenuation power-law exponent, and $d_{0}$ is
sufficiently large that the signals are in the far field. It is assumed that
no mobiles are within distance $d_{0}$ of any base station.

It is assumed that the \{$g_{i,j}\}$ remain fixed for the duration of a time
interval, but vary independently from interval to interval (block fading).
With probability $p_{i}$, the $i^{th}$ interferer transmits in the same time
interval as the reference signal. The \emph{activity probability} $p_{i}$ can
be used to model voice-activity factors or controlled silence. Although the
$\{p_{i}\}$ need not be the same, it is assumed that they are identical in the
subsequent examples.

Let $\mathsf{g}(i)$ denote a function that returns the index of the sector
antenna serving $X_{i}$ so that $X_{i}\in{\mathcal{X}}_{j}$ if $\mathsf{g}%
\left(  i\right)  =j$. Usually, the sector antenna $S_{\mathbf{g}\left(
i\right)  }$ that serves mobile $X_{i}$ is selected to be the one with index
\begin{equation}
\mathsf{g}\left(  i\right)  =\underset{j}{\mathrm{argmax}}\,\left\{
10^{\xi_{i,j}/10}f\left(  ||S_{j}-X_{i}||\right)  ,\text{ }X_{i}\in
\mathcal{A}_{j}\right\}
\end{equation}
which is the sector antenna with minimum path loss from $X_{i}$ among those
that cover $X_{i}$. In the absence of shadowing, it will be the sector antenna
that is closest to $X_{i}$. In the presence of shadowing, a mobile may
actually be associated with a sector antenna that is more distant than the
closest one if the shadowing conditions are sufficiently better.

The instantaneous SINR at sector antenna $S_{j}$ when the desired signal is
from $X_{r}\in\mathcal{X}_{j}$ is%

\begin{equation}
\gamma_{r,j}=\frac{\rho_{r,j}}{\displaystyle{\mathcal{N}}+\sum_{i=1,i\neq
r}^{M}I_{i}\rho_{i,j}} \label{eqn:SINR1}%
\end{equation}
where $\mathcal{N}$ is the noise power, M is the number of mobiles, and
$I_{i}$ is is a Bernoulli variable with probability $P[I_{i}=1]=p_{i}$ and
$P[I_{i}=0]=1-p_{i}$. Substituting (\ref{eqn:power}) and (\ref{eqn:pathloss})
into (\ref{eqn:SINR1}) yields
\begin{equation}
\gamma_{r,j}=\frac{g_{r,j}\Omega_{r,j}}{\displaystyle\Gamma^{-1}%
+\sum_{i=1,i\neq r}^{M}I_{i}g_{i,j}\Omega_{i,j}}%
\end{equation}
where $\Gamma=d_{0}^{\alpha}P_{r}/\mathcal{N}$ is the signal-to-noise ratio
(SNR) due to a mobile located at unit distance when fading and shadowing are
absent, and
\begin{eqnarray}
\Omega_{i,j}
&  = &
\begin{cases}
10^{\xi_{r,j}/10}||S_{j}-X_{r}||^{-\alpha} & i=r\\
\displaystyle\frac{hP_{i}}{GP_{r}}10^{\xi_{i,j}/10}||S_{j}-X_{i}||^{-\alpha} &
i:X_{i}\in\mathcal{A}_{j}\backslash X_{r}\\
0 & i:X_{i}\notin\mathcal{A}_{j}%
\end{cases}
\nonumber \\
\label{omega}%
\end{eqnarray}
is the normalized mean despread power of $X_{i}$ received at $S_{j}$, where
the normalization is by $P_{r}$. The set of $\{\Omega_{i,j}\}$ for reference
receiver $S_{j}$ is represented by $\boldsymbol{\Omega}%
_{j}=\{\Omega_{1,j},...,\Omega_{M,j}\}$.

\section{Outage Probability}

\label{Section:Outage}

Let $\beta_{r}$ denote the minimum instantaneous SINR required for reliable
reception of a signal from $X_{r}$ at its serving sector antenna.
An \emph{outage} occurs when the SINR of a signal
from $X_{r} $ falls below $\beta_{r}$. The value of $\beta_{r}$ is a function
of the \emph{rate} $R_{r}$ of the uplink.
The relationship $R_{r}=C(\beta_{r})$ depends on the
modulation and coding schemes used, and typically only a discrete set of
$R_{r}$ can be selected. While the exact dependence of $R_{r}$ on $\beta_{r}$
can be determined empirically through tests or simulation, we make the
simplifying assumption when computing our numerical results that $C(\beta
_{r})=\log_{2}(1+\beta_{r})$ corresponding to the Shannon capacity for complex
discrete-time AWGN channels. This assumption is fairly accurate for modern
cellular systems,
which use turbo codes with a large number of available rates.

Conditioning on $\boldsymbol{\Omega}_{j}$, the outage probability of a desired
signal from $X_{r}\in{\mathcal{X}}_{j}$ that arrives at $S_{j}, j=\mathbf{g}(r),$
is
\begin{equation}
\epsilon_{r}=P\left[  \gamma_{r,j}\leq\beta_{r}\big|\boldsymbol{\Omega}%
_{j}\right]  .
\end{equation}
Because it is conditioned on $\boldsymbol{\Omega}_{j}$, the outage probability
depends on the particular network realization, which has dynamics over
timescales that are much slower than the fading. In \cite{tor2}, it is proved
that
\begin{equation}
\epsilon_{r}=1-e^{-\beta_{0}z}\sum_{s=0}^{m_{0}-1}{\left(  \beta_{0}z\right)
}^{s}\sum_{t=0}^{s}\frac{z^{-t}H_{t}(\boldsymbol{\Psi})}{(s-t)!} \label{a1}%
\end{equation}
where $m_{0}=m_{r,j}$ is an integer, $\beta_{0}=\beta_{r}m_{0}/\Omega_{r,j}$,
$z=\Gamma^{-1},$
\begin{align}
\Psi_{i}  &  =\left(  \beta_{0}\frac{\Omega_{i,j}}{m_{i,j}}+1\right)
^{-1}\hspace{-0.5cm},\hspace{1cm}\mbox{for $i=\{1,...,M\}$, }\\
H_{t}(\boldsymbol{\Psi})  &  =\mathop{ \sum_{\ell_i \geq 0}}_{\sum_{i=0}%
^{M}\ell_{i}=t}\prod_{i=1,i\neq r}^{M}G_{\ell_{i}}(\Psi_{i}),
\end{align}%
\begin{equation}
G_{\ell}(\Psi_{i})=%
\begin{cases}
1-p_{i}(1-\Psi_{i}^{m_{i,j}}) & \mbox{for $\ell=0$}\\
\frac{p_{i}\Gamma(\ell+m_{i,j})}{\ell!\Gamma(m_{i,j})}\left(  \frac
{\Omega_{i,j}}{m_{i,j}}\right)  ^{\ell}\Psi_{i}^{m_{i,j}+\ell} & \mbox{for
$\ell>0$.}
\end{cases}
\label{a2}%
\end{equation}
In (\ref{a2}), the index $i$ on the right hand side matches that of the argument $\Psi_{i}$,
and $j=\mathbf{g}(r)$ to coincide with the index of the serving sector antenna.

\section{Network Policies\label{Section:Policies}}

\subsection{Power Control}

\label{pc} A typical power-allocation policy for DS-CDMA networks is to select
the transmit power $\{P_{i}\}$ for all mobiles in $\mathcal{X}_{j}$
such that, after compensation for shadowing and power-law attenuation, each
mobile's transmission is received at sector antenna $S_{j}$ with the same
power $P_{0}$.
For such a power-control policy, each mobile in $\mathcal{X}_{j}$ will
transmit with a power $P_{i}$ that satisfies
\begin{equation}
{P}_{i}10^{\xi_{i,j}/10}f\left(  ||S_{j}-X_{i}||\right)  =P_{0},\text{
\ }X_{i}\in\mathcal{X}_{j}. \label{Eqn:Intracell}%
\end{equation}
where $f(\cdot)$ is given by (\ref{eqn:pathloss}).
Since the reference mobile $X_{r}\in\mathcal{X}_{j}$, its transmit power is
determined by (\ref{Eqn:Intracell}).
To accomplish the power-control policy,
sector-antenna receivers estimate the average received powers of their
associated mobiles. Feedback of these estimates enables the associated mobiles
to change their transmitted powers so that all received powers are
approximately equal \cite{zhao}.

For a reference mobile $X_{r}$, the interference at sector antenna $S_{j}$ is
from the mobiles in the set $\mathcal{A}_{j}\backslash X_{r}$. This set can be
partitioned into two subsets. The first subset $\mathcal{X}_{j}\backslash
X_{r}$ comprises the \textit{intracell interferers}, which are the other
mobiles in the same cell and sector as the reference mobile. The second subset
$\mathcal{A}_{j}\backslash\mathcal{X}_{j}$ comprises the \textit{intercell
interferers}, which are the mobiles covered by sector antenna $S_{j}$ but
associated with a cell sector other than $\mathcal{X}_{j} $.

Considering intracell interference, all of the mobiles within the sector
transmit with power given by (\ref{Eqn:Intracell}).   Since $P_r$ and $P_i$
are obtained from (\ref{Eqn:Intracell}), the middle
line of (\ref{omega}) gives the normalized mean received power of the intracell interferers:
\begin{equation}
\Omega_{i,j}=\frac{h}{G}10^{\xi_{r,j}/10}||S_{j}-X_{r}||^{-\alpha},\text{
\ }X_{i}\in\mathcal{X}_{j}\backslash X_{r}. \label{a3}%
\end{equation}
Although the number of mobiles $M_{j}$ in the cell sector must be known to
compute the outage probability, the locations of these mobiles in the cell are
irrelevant to the computation of the $\Omega_{i,j}$ of the intracell interferers.

Considering intercell interference, the set $\mathcal{A}_{j}\backslash
\mathcal{X}_{j}$ can be further partitioned into sets $\mathcal{A}_{j}%
\cap\mathcal{X}_{k}$, $k\neq j$, containing the mobiles covered by sector
antenna $S_{j}$ but associated with some other sector antenna $S_{k}$. For
those mobiles in $\mathcal{A}_{j}\cap\mathcal{X}_{k}$, power control implies
that
\begin{equation}
{P}_{i}10^{\xi_{i,k}/10}f\left(  ||S_{k}-X_{i}||\right)  =P_{0},\text{
\ }X_{i}\in\mathcal{X}_{k}\cap A_{j},\text{ \ }k\neq j. \label{Eqn:Inter}%
\end{equation}
Substituting (\ref{Eqn:Inter}), $\left(  \ref{Eqn:Intracell}\right)  $ with
$i=r$, and (\ref{eqn:pathloss}) into $\left(  \ref{omega}\right)  $ yields%

\begin{align}
\Omega_{i,j}  &  =\frac{h}{G}10^{\xi_{i,j}^{\prime}/10}\left(  \frac
{||S_{j}-X_{i}||||S_{j}-X_{r}||}{||S_{k}-X_{i}||}\right)  ^{-\alpha}\text{
\ }\nonumber\\
\xi_{i,j}^{\prime}  &  =\xi_{i,j}+\xi_{r,j}-\xi_{i,k},\text{ \ }X_{i}%
\in\mathcal{X}_{k}\cap A_{j},\text{ \ }k\neq j\text{ \ } \label{a4}%
\end{align}
for $\mathcal{A}_{j}\backslash\mathcal{X}_{j}$, which gives the normalized mean intercell
interference power at the reference sector antenna due to interference from
mobile $i$ of sector $k=\mathsf{g}(i)$.

\subsection{Rate Control}

\label{Section:RateControl}

In addition to controlling the transmitted power, the rate $R_{i}$ of each
uplink needs to be selected. Due to the irregular network geometry, which
results in cell sectors of variable areas and numbers of mobiles,
the amount of received interference can vary dramatically
from one sector antenna to another. With a fixed rate, or equivalently, a fixed SINR
threshold $\beta$ for each sector, the result is a highly variable outage
probability. An alternative to using a fixed rate for the entire network is to
adapt the rate of each uplink to satisfy an outage constraint or maximize the
throughput of each uplink \cite{chai}, \cite{sub}.

To illustrate the influence of rate on performance, consider the following
example. The network has $C=50$ base stations and $M=400$ mobiles placed in a
circular network of radius $r_{\mathsf{net}}=2$. The base-station exclusion
zones have radius $r_{\mathsf{bs}}=0.25$, and the mobile exclusion zones
have radius $r_{\mathsf{m}}=0.01$. The spreading factor is $G=16,$ and the
chip factor is $h=2/3$. Since $M/C=G/2$, the network is characterized as being
\emph{half loaded}. The SNR is $\Gamma=10$ dB, the activity factor is
$p_{i}=1$, the path-loss exponent is $\alpha=3$, and lognormal shadowing is
assumed with standard deviation $\sigma_{s}=8$ dB. A \emph{distance-dependent
fading} model is assumed, where the Nakagami parameter $m_{i,j}$ is
\begin{equation}
m_{i,j}=%
\begin{cases}
3 & \mbox{ if }\;||S_{j}-X_{i}||\leq r_{\mathsf{bs}}/2\\
2 & \mbox{ if }\;r_{\mathsf{bs}}/2<||S_{j}-X_{i}||\leq r_{\mathsf{bs}}\\
1 & \mbox{ if }\;||S_{j}-X_{i}||>r_{\mathsf{bs}}%
\end{cases}
.
\label{eqn:distance_dependent}
\end{equation}
The distance-dependent-fading model characterizes the situations where mobiles
close to the base station are in the line-of-sight, but mobiles farther away
are not.

\begin{figure}[tb]%
\centering
\includegraphics[width=9cm]{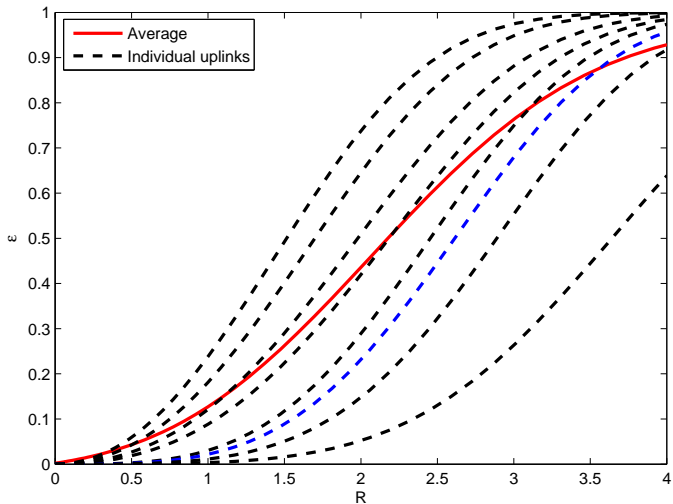}
\caption{Outage probability of eight randomly selected uplinks (dashed lines)
along with the average outage probability for the entire network (solid line).
The results are for a half-loaded network ($M/C=G/2$) with distance-dependent
fading and shadowing ($\sigma_{s}=8$ dB) and are shown as a function of the
rate $R$. }
\label{Figure:OutagePC}
\vspace{-0.5cm}
\end{figure}

Fig. \ref{Figure:OutagePC} shows the outage probability as a function of rate.
Assuming the use of a capacity-approaching code, two-dimensional signaling
over an AWGN channel, and Gaussian interference, the SINR threshold
corresponding to rate $R$ is $\beta=2^{R}-1$. The dashed lines in Fig.
\ref{Figure:OutagePC} were generated by selecting eight random uplinks and
computing the outage probability for each using this threshold. Despite the
power control, there is considerable variability in the outage
probability. The outage probabilities \{$\epsilon_{i}\}$ were computed for all
$M$ uplinks in the system, and the average outage probability,
\begin{equation}
\mathbb{E}[\epsilon]=\frac{1}{M}\sum_{i=1}^{M}\epsilon_{i} \label{a5}%
\end{equation}
is displayed as a solid line in the figure.%

\begin{figure}[tb]%
\centering
\includegraphics[width=9cm]{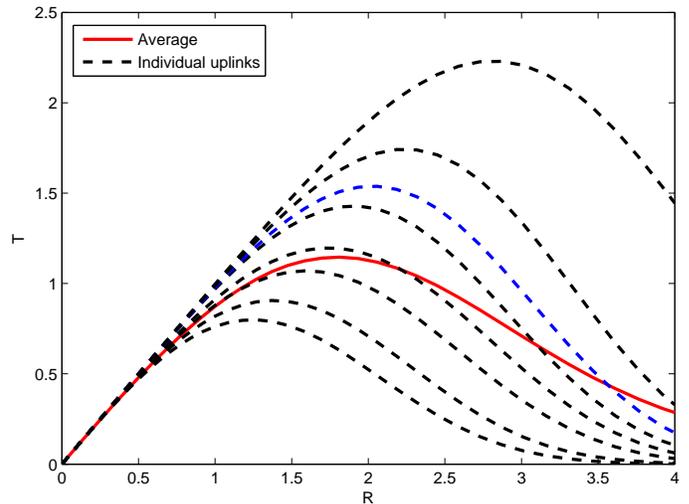}
\caption{Throughput of eight randomly selected uplinks (dashed lines) along
with the average throughput for the entire network (solid line). System
parameters are the same used to generate Fig. \ref{Figure:OutagePC}.}
\label{Figure:TCPC}
\vspace{0.2cm}
\end{figure}

Fig. \ref{Figure:TCPC} shows the \emph{throughput} as a function of rate,
where the throughput of the $i^{th}$ uplink is found as
\begin{equation}
T_{i}=R_{i}(1-\epsilon_{i}) \label{a6}%
\end{equation}
and represents the rate of successful transmissions. The parameters are the
same as those used to produce Fig. \ref{Figure:OutagePC}, and again the SINR
threshold corresponding to rate $R$ is $\beta=2^{R}-1$. The plot shows the
throughput for the same eight uplinks whose outages were shown in Fig.
\ref{Figure:OutagePC}, as well as the average throughput
\begin{equation}
\mathbb{E}[T]=\frac{1}{M}\sum_{i=1}^{M}R_{i}(1-\epsilon_{i}). \label{a7}%
\end{equation}

\subsubsection{Fixed-Rate Policies}

A fixed-rate policy requires that all uplinks in the system must use the same
rate; i.e., $R_{i}=R$ for all uplinks. On the one hand, the rate could be
selected to maximize the average throughput. With respect to the example shown
in Fig. \ref{Figure:TCPC}, this corresponds to selecting the $R$ that
maximizes the solid curve, which occurs at $R=1.81$. However, at the rate that
maximizes throughput, the corresponding outage probability could be
unacceptably high. When $R=1.81$ in the example, the corresponding average
outage probability is $\mathbb{E}[\epsilon]=0.37$, which is too high for many
applications. As an alternative to maximizing throughput, the rate $R$ could
be selected to satisfy an outage constraint $\zeta$ so that $\mathbb{E}%
[\epsilon]\leq\zeta$. For instance, setting $R=0.84$ in the example satisfies
an average outage constraint $\zeta=0.1$ with equality. To distinguish between
the two fixed-rate policies, we call the first policy \emph{maximal-throughput
fixed rate} (MTFR) and the second policy \emph{outage-constrained fixed rate} (OCFR).

\subsubsection{Variable-Rate Policies}

If $R$ is selected to satisfy an average outage constraint, the outage
probability of the individual uplinks will vary around this average.
Furthermore, selecting $R$ to maximize the average throughput does not
generally maximize the throughput of the individual uplinks. These issues can
be alleviated by selecting the rates $R_{i}$ independently for the different
uplinks. The rates could be selected to require all uplinks to satisfy the
outage constraint $\zeta$; i.e., $\epsilon_{i}\leq\zeta$ for all $i$. We call
this the \emph{outage-constrained variable-rate} (OCVR) policy. Alternatively,
the selection could be made to maximize the throughput of each uplink; i.e.,
$R_{i}=\arg\max T_{i}$ for each uplink, where the maximization is over all
possible rates. We call this the \emph{maximal-throughput variable-rate}
(MTVR) policy. Both policies can be implemented by having the base station
track the outage probabilities or throughputs of each uplink and feeding back
rate-control commands to ensure that the target performance is achieved.
The outage probability can be easily found by encoding the data with a
cyclic-redundancy-check code and declaring an outage when a check fails.

For both variable-rate policies, we assume
that the code rate is adapted by maintaining the duration of channel symbols
while varying the number of information bits per symbol. The spreading factor
$G$ and symbol rate are held constant, so there is no change in bandwidth.
However, a major drawback with rate control is that the rates required to
maintain a specified outage probability varies significantly among the mobiles
in the network. This variation results in low throughput for some mobiles,
particularly those located at the edges of the cells, while other mobiles have
a high throughput. Unequal throughputs may not be acceptable when a mobile may
be stuck or parked near a cell edge for a long time, and an interior mobile
may have more allocated throughput than it needs.


\subsection{Cell Association}

The downlinks of a cellular DS-CDMA network use orthogonal spreading
sequences. Because the number of orthogonal spreading sequences available to
the cell sector is limited to $G$ when sectorization is used for the
downlinks, the number of served mobiles using $S_{j}$ is limited to $M_{j}\leq
G$. This limit then restricts the number of mobiles that have uplink service.
If there are $M_{j}>G$ mobiles, then some of these mobiles will either be
refused service by $S_{j}$ or given service at a lower rate (through the use
of an additional time multiplexing). In the following, we consider two
policies for handling this situation. With the first policy, which we call
\emph{denial policy}, the $M_{j}-G$ mobiles whose path losses to the
base station are greatest are denied service, in which case they do not appear
in the set ${\mathcal{X}}_{j}$ for any $j$, and their rates are set to zero.
With the second policy, which we call \emph{reselection policy}, each
of the $M_{j}-G$ mobiles in an overloaded cell sector attempts to connect to
the sector antenna with the next-lowest path loss out to a maximum
reassociation distance $d_{\mathsf{max}}$. If no suitably oriented sector
antenna is available within distance $d_{\mathsf{max}}$, the mobile is denied
service. The \emph{denial} policy is the same as the \emph{reselection} policy
with $d_{\mathsf{max}}=0$.
\section{Performance Analysis}

\label{Section:Performance}

\subsection{Performance Metrics}

While the outage probability, throughput, and rate characterize the
performance of a single uplink, they do not quantify the total data flow in
the network because they do not account for the number of uplink users that
are served. By taking into account the number of mobiles per unit area, the
total data flow in a given area can be characterized by the \emph{area
spectral efficiency}, defined as
\begin{equation}
\mathcal{A}=\lambda\mathbb{E}[T]=\lambda\mathbb{E}\left[  \left(
1-\epsilon\right)  R\right]  \label{eqn:tc}%
\end{equation}
where $\lambda=M/(\pi r_{\mathsf{net}}^{2})$ is the density of transmissions in
the network, and the units are bits per channel use per unit area. Area
spectral efficiency \cite{web} can be interpreted as the spatial intensity of
transmissions; i.e., the rate of successful data transmission per unit area.

Performance metrics are calculated by using a Monte Carlo approach with 1000
simulation trials as follows. In each simulation trial, a realization of the
network is obtained by placing $C$ base stations and $M$ mobiles within the
disk of radius $r_{\mathsf{net}}$ according to the uniform-clustering model
with minimum base-station separation $r_{\mathsf{bs}}$ and minimum mobile
separation $r_{\mathsf{m}}$. The path loss from each base station to each
mobile is computed by applying randomly generated shadowing factors. The set
of mobiles associated with each cell sector is determined. Assuming that the
number of mobiles served in a cell sector cannot exceed $G$, which is the
number of orthogonal sequences available for the downlink, the rate of the
last $M_{j}-G$ mobiles in the cell sector (if there are that many) is
determined by the cell association policy.  At each sector antenna, the
power-control policy is applied to determine the power the antenna receives
from each mobile that it serves. In each cell sector, the rate-control policy
is applied to determine the rate and threshold.

For each uplink, the outage probability is computed by applying the
rate-control policy and using $\left(  \ref{a1}\right)  -\left(
\ref{a2}\right)  $, $\left(  \ref{a3}\right)  $, and $\left(  \ref{a4}\right)
$.  Equation $\left(  \ref{a5}\right)  $\ is applied to calculate
the average outage probability for the network realization. Equation $\left(
\ref{a6}\right)  $\ is applied to calculate the throughputs according to the
resource-allocation policy, and then $\left(  \ref{a7}\right)
$\ is applied to compute the average throughput for the network
realization. Finally, multiplying by the density $\lambda$ and averaging over all simulated network realizations yields the average
outage probability $\overline{\epsilon}$\ and the average area spectral
efficiency $\overline{\mathcal{A}}.$

Notice that the simulator only needs to randomly place the mobiles and base stations according to the spatial model and select the shadowing factors; i.e., it only needs to draw the {\em network realization}.  The simulator does {\em not} need to draw the power gains $\{g_{i,j}\}$.  Without applying the results of Section \ref{Section:Outage}, the simulator would be far more complex because it would need to simulate the fading for every network realization.    The number of power gains that would need to be generated could be as many as a million for every network realization, depending on the desired level of confidence.  On the other hand, our approach not only avoids the unnecessary generation of power gains, it provides an {\em exact} result for any given network realization.  In essence, the approach makes it possible to achieve the same results as a very detailed simulator, but can do so without needing to simulate the fading.

\subsection{Simulation Parameters}

In the following subsections, the Monte Carlo method described in the previous
subsection is used to characterize the uplink performance. In all cases
considered, the network has $C=50$ base stations placed in a circular network
of radius $r_{\mathsf{net}}=2$. Except for Subsection \ref{Section:RBS}, which
studies the influence of $r_{\mathsf{bs}}$, the base-station exclusion zones
are set to have radius $r_{\mathsf{bs}}=0.25$. A variable number $M$ of
mobiles are placed within the network using exclusion zones of radius
$r_{\mathsf{m}}=0.01$. The SNR is $\Gamma=10$ dB, and the activity factor is
$p_{i}=1$. Unless otherwise stated, the path-loss exponent is $\alpha=3$. Two
fading models are considered: \emph{Rayleigh fading}, where $m_{i,j}=1$ for
all $i$, and \emph{distance-dependent fading}, which is described by
(\ref{eqn:distance_dependent}). Both unshadowed and shadowed
($\sigma_{s}=8$ dB) environments are considered. The chip factor is $h=2/3$,
and except for Subsection \ref{Section:PG}, which studies the influence of
$G$, the spreading factor is $G=16$. Except for Section
\ref{Section:Selection}, $d_{\mathsf{max}}=0$; i.e., mobiles that overload
sectors are denied service.

\subsection{Outage-Constrained Fixed-Rate Policy}

\begin{figure}[tb]%
\centering
\includegraphics[width=9cm]{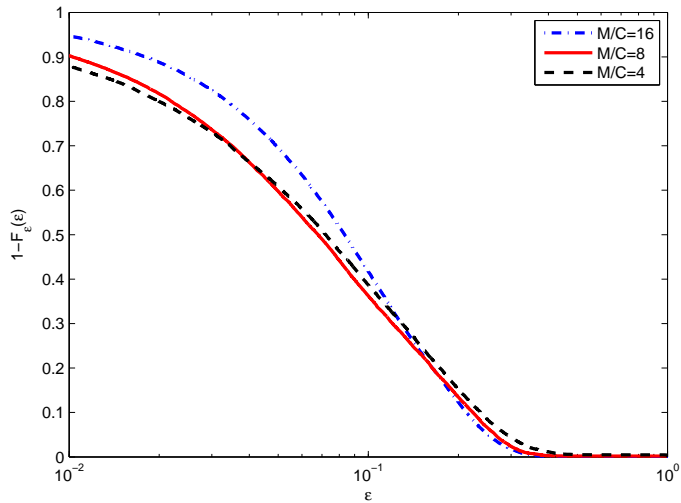}
\caption{Ccdf of the outage probability using an OCFR policy, $R=2$, three
network loads, distance-dependent fading, and shadowing. }
\label{Figure:CcdfPC}
\vspace{-0.15cm}
\end{figure}

As illustrated in Fig. \ref{Figure:OutagePC}, using an OCFR policy results in
a high variability of outage probabilities. Fig. \ref{Figure:CcdfPC}
illustrates the variability of $\epsilon$ with respect to all uplinks and
simulation trials under a OCFR policy by plotting its complementary cumulative
distribution function (ccdf) with $R=2$, three
network loads, distance-dependent fading, and shadowing.%

\subsection{Outage-Constrained Variable-Rate Policy}

\begin{figure}[tb]
\centering
\includegraphics[width=9cm]{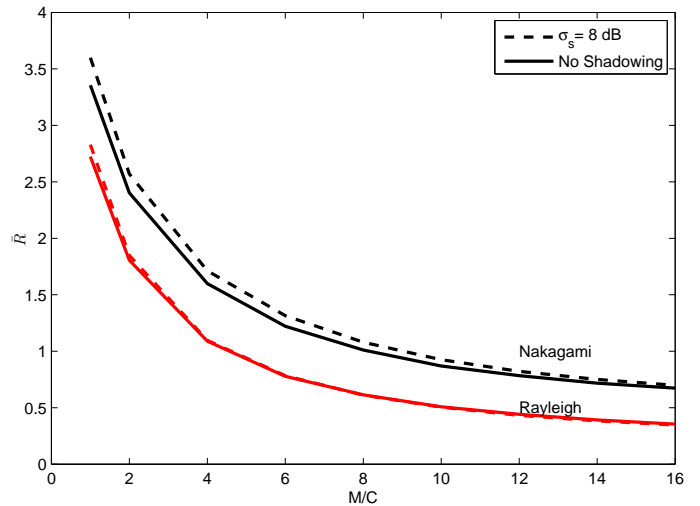}
\caption{Average rate of the OCVR policy as function of the load $M/C$ for
both Rayleigh and distance-dependent Nakagami fading, and both shadowed
($\sigma_{s}=$ 8 dB) and unshadowed cases. }
\label{Figure:Rate}
\end{figure}

\begin{figure}[tb]%
\centering
\includegraphics[width=8.75cm]{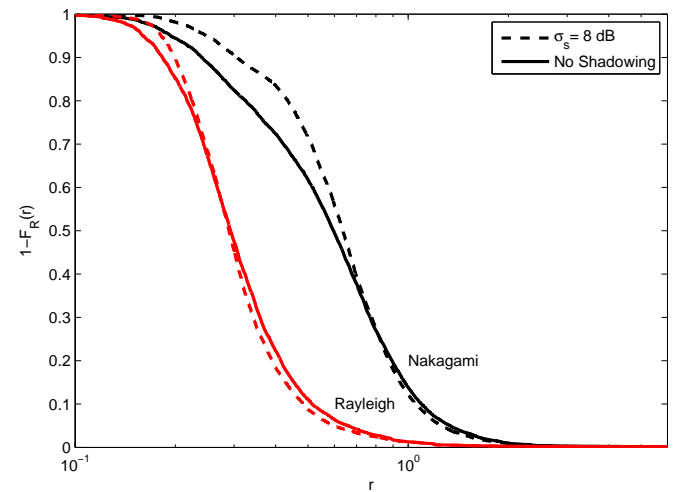}
\caption{Ccdf of the rate for fully-loaded system ($M/C=G$) under the OCVR
policy in Rayleigh and distance-dependent Nakagami fading, and both shadowed
($\sigma_{s}=$ 8 dB) and unshadowed cases.}
\label{Figure:CcdfRate}
\vspace{-0.7cm}
\end{figure}

With the OCVR policy, the rate $R_{i}$ (equivalently, $\beta_{i}$) of each
uplink is selected such that the outage probability does not exceed
$\zeta=0.1$. While the outage probability is fixed, the rates of the uplinks
will be variable. Let $\overline{R}$ denote the average rate over all uplinks
and simulation trials. In Fig. \ref{Figure:Rate}, $\overline{R}$ \ is shown as
a function of the load $M/C.$ In Fig. \ref{Figure:CcdfRate}, the variability
of $R_{i}$ is illustrated by showing the ccdf of the rate for a fully-loaded
system ($M/C=G=16$) in both Rayleigh fading and distance-dependent fading, and
both with and without shadowing. The fairness of the system can be determined
from this figure, which shows the percentage of uplinks that meet a particular
rate requirement.

\begin{figure}[tb]%
\centering
\includegraphics[width=8.75cm]{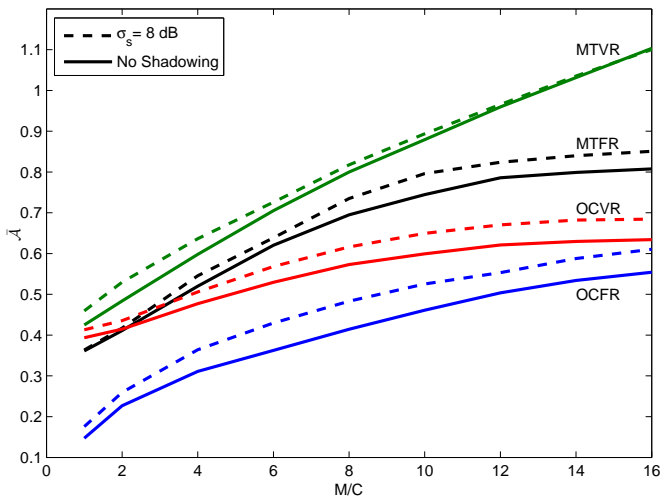} \vspace{-0.45cm}
\caption{ Average area spectral efficiency for the four network policies as
function of the load $M/C$ for distance-dependent fading and both shadowed
($\sigma_{s}=$ 8 dB) and unshadowed cases. }
\label{Figure:TCComp}
\vspace{0.05cm}
\end{figure}

\subsection{Policy Comparison}

Fig. \ref{Figure:TCComp} shows the average area spectral efficiency
$\overline{\mathcal{A}}$ of the four network policies in distance-dependent
fading, both with and without shadowing, as a function of the load $M/C$. For
the OCFR policy, the optimal rate was determined for each simulation trial and then $\bar{\mathcal A}$ was computed by averaging over 1000 trials.
For the OCVR policy, the
uplink rate $R_{i}$ of each uplink was maximized subject to an outage
constraint. While the area spectral efficiencies of the MTFR and MTVR policies
are potentially superior to those of the OCFR and OCVR policies, this
advantage comes at the cost of a variable and high value of $\epsilon$, which
is generally too large for most applications. The
OCVR\ policy has a higher average
area spectral efficiency than the OCFR policy.

\subsection{ Spreading factor}

\label{Section:PG}

\begin{figure}[ptb]%
\includegraphics[width=9cm]{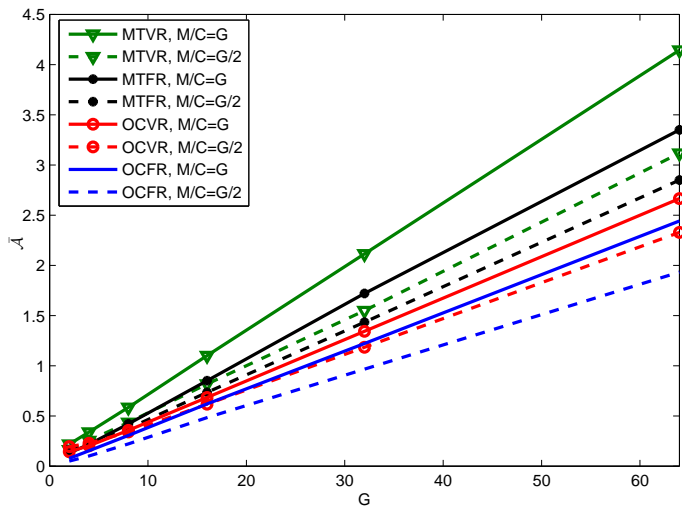}
\caption{Average area spectral efficiency as function of spreading factor $G$
for two values of system load, distance-dependent fading, and shadowing with
$\sigma_{s}=8$ dB. }
\label{Figure:PG}
\vspace{-0.1cm}
\end{figure}

Fig. \ref{Figure:PG} shows $\overline{\mathcal{A}}$ as a function of the
spreading factor $G$ (with $h=2/3$) for the shadowed distance-dependent-fading
channel. Two loads are shown for each of the four policies. An increase in $G$
is beneficial for all policies, but the MTVR policy benefits the most.%

\subsection{ Base-station Exclusion Zone}

\label{Section:RBS}

\begin{figure}[tb]%
\centering
\vspace{-0.1cm}
\includegraphics[width=9cm]{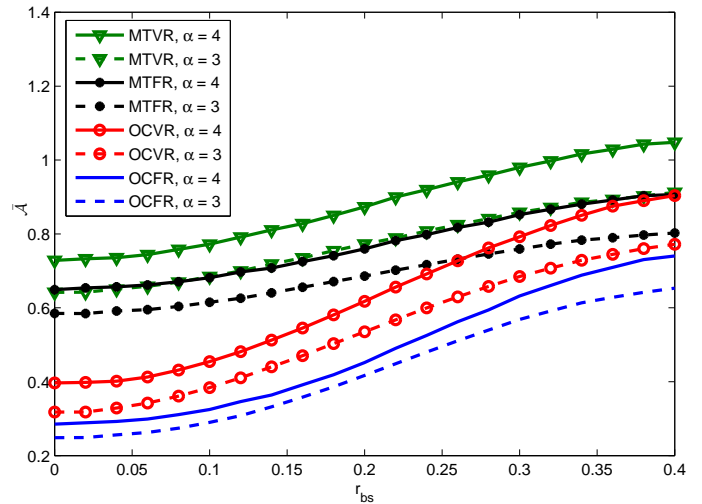} \vspace{-0.55cm}
\caption{Average area spectral efficiency as a function of the base-station
exclusion-zone radius $r_{\mathsf{bs}}$ for four policies and two values of
path-loss exponent $\alpha$.}
\label{Figure:TCRbs}
\end{figure}

Fig. \ref{Figure:TCRbs} shows $\overline{\mathcal{A}}$ for each of the four
policies as a function of the base-station exclusion-zone radius
$r_{\mathsf{bs}}$ for $M/C=G/2$ and two values of path-loss exponent $\alpha$.
The distance-dependent fading model is used, and shadowing is applied with
$\sigma_{s}=$ 8 dB. The two policies that constrain the outage probability are
more sensitive to the value of $r_{\mathsf{bs}}$ than the two policies that
maximize throughput. An increase in $\alpha$ increases $\bar{\mathcal A}$ for all four policies.

\begin{figure}[tb]%
\centering
\includegraphics[width=9cm]{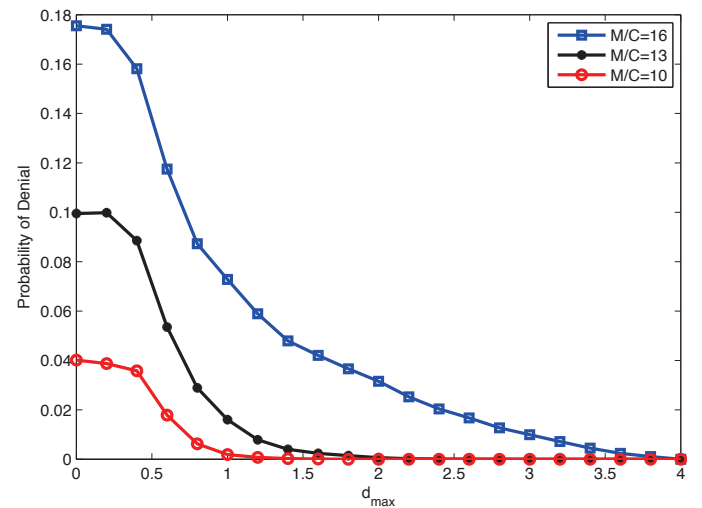}
\caption{Probability that a mobile is denied service due to cell overload as a
function of the maximum reselection distance $d_{\mathsf{max}}.$ }%
\label{Figure:Reselection}
\vspace{0.1 cm}
\end{figure}

\subsection{ Cell Association}

\label{Section:Selection}

\begin{figure}[tb]%
\centering
\includegraphics[width=9cm]{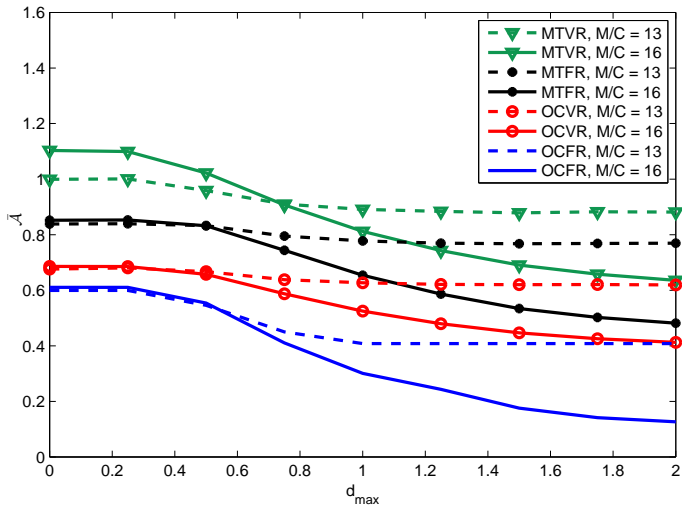}

\caption{Average area spectral efficiency as a function of the maximum
reselection distance $d_{\mathsf{max}}.$ }
\label{Figure:TCdmax}

\end{figure}

In the previously shown results, mobiles were denied service if they were in
an overloaded sector. By allowing mobiles to reselect base stations out to a
distance of $d_{\mathsf{max}}$, the probability that they are denied service
due to overload can be reduced. Fig. \ref{Figure:Reselection} shows the
probability that a mobile is denied service due to sector overload as a
function of the maximum reselection distance $d_{\mathsf{max}}$ for three
different values of average cell load $M/C$. By setting $d_{\mathsf{max}}$
sufficiently large, the probability that a mobile is denied service due to
overload can be made arbitrarily small. The tradeoff for keeping these mobiles
connected is that $\overline{\mathcal{A}}$ decreases because mobiles must
transmit at a higher power, thereby causing more interference. This tradeoff
is illustrated in Fig. \ref{Figure:TCdmax}, which shows $\overline
{\mathcal{A}}$ as a function of $d_{\mathsf{max}}$ for all four policies and
two system loads. In general, $\overline{\mathcal{A}}$ decreases with
increasing $d_{\mathsf{max}}$, and the decrease is more pronounced for the
more heavily loaded system.%

\section{Conclusion}

A new analysis of DS-CDMA uplinks has been presented. This analysis is much
more detailed and accurate than existing ones and facilitates the resolution
of network design issues. In particular, it has been shown that once power
control is established, the rate can be allocated according to a fixed-rate or
variable-rate policy with the objective of either maximizing throughput or
meeting an outage constraint. An advantage of variable-rate power control is
that it allows an outage constraint to be enforced on every uplink, which is
impossible when a fixed rate is used throughout the network. Another advantage
is an increased area spectral efficiency.


\balance

\ifpdf
  \begin{IEEEbiography}{Don Torrieri}
\else
  \begin{IEEEbiography}[{\includegraphics[width=1in,height=1.25in,clip,keepaspectratio]{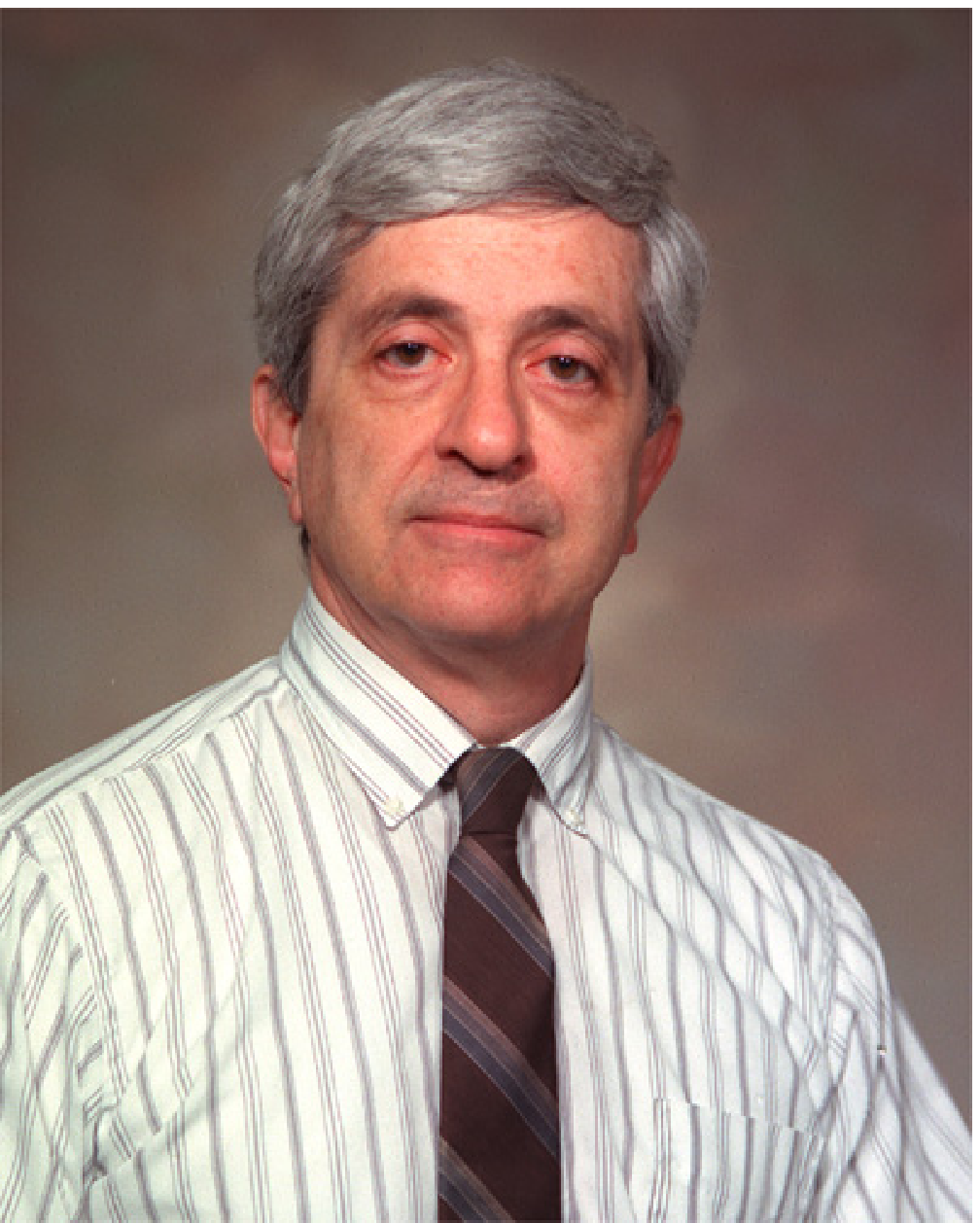}}]{Don Torrieri}
\fi
 is a research engineer and Fellow of the US Army Research Laboratory. His primary research interests are communication systems, adaptive arrays, and signal processing. He received the Ph. D. degree from the University of Maryland. He is the author of many articles and several books including {\em Principles of Spread-Spectrum Communication Systems}, 2nd ed. (Springer, 2011). He has taught many graduate courses at Johns Hopkins University and many short courses. In 2004, he received the Military Communications Conference achievement award for sustained contributions to the field.
\end{IEEEbiography}

\ifpdf
  \begin{IEEEbiography}{Matthew C. Valenti}
\else
  \begin{IEEEbiography}[{\includegraphics[width=1in,height=1.25in,clip,keepaspectratio]{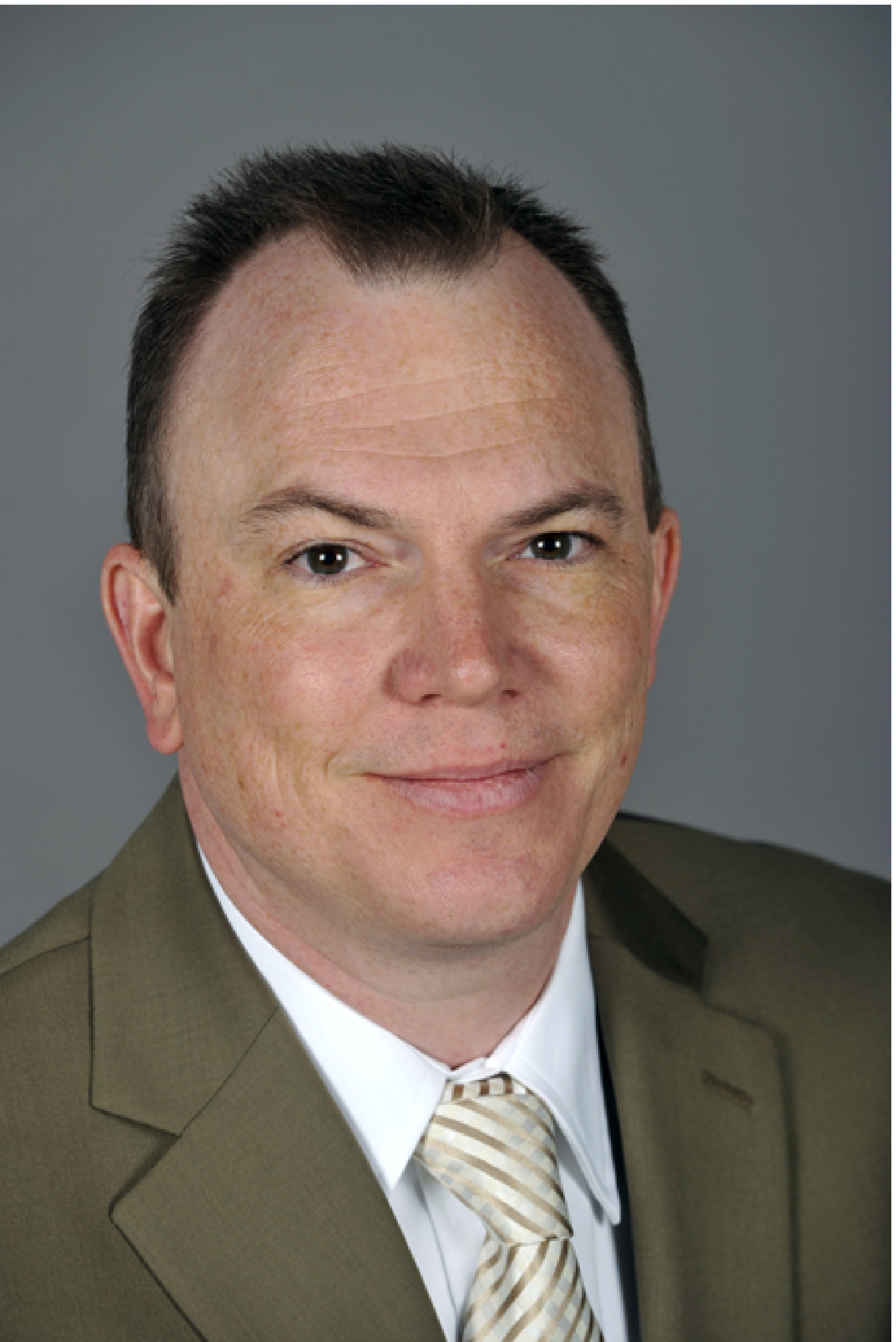}}]{Matthew C. Valenti}
\fi
is a Professor in Lane Department of Computer Science and Electrical Engineering at West Virginia University. He holds BS and Ph.D. degrees in Electrical Engineering from Virginia Tech and a MS in Electrical Engineering from the Johns Hopkins University. From 1992 to 1995 he was an electronics engineer at the US Naval Research Laboratory.  He serves as an associate editor for {\em IEEE Wireless Communications Letters} and as Vice Chair of the Technical Program Committee for Globecom-2013.  Previously, he has served as a track or symposium co-chair for VTC-Fall-2007, ICC-2009, Milcom-2010, ICC-2011, and Milcom-2012, and has served as an editor for {\em IEEE Transactions on Wireless Communications} and {\em IEEE Transactions on Vehicular Technology}. His research interests are in the areas of communication theory, error correction coding, applied information theory, wireless networks, simulation, and secure high-performance computing.  His research is funded by the NSF and DoD.  He is registered as a Professional Engineer in the State of West Virginia.
\end{IEEEbiography}

\ifpdf
  \begin{IEEEbiography}{Salvatore Talarico}
\else
  \begin{IEEEbiography}[{\includegraphics[width=1in,height=1.25in,clip,keepaspectratio]{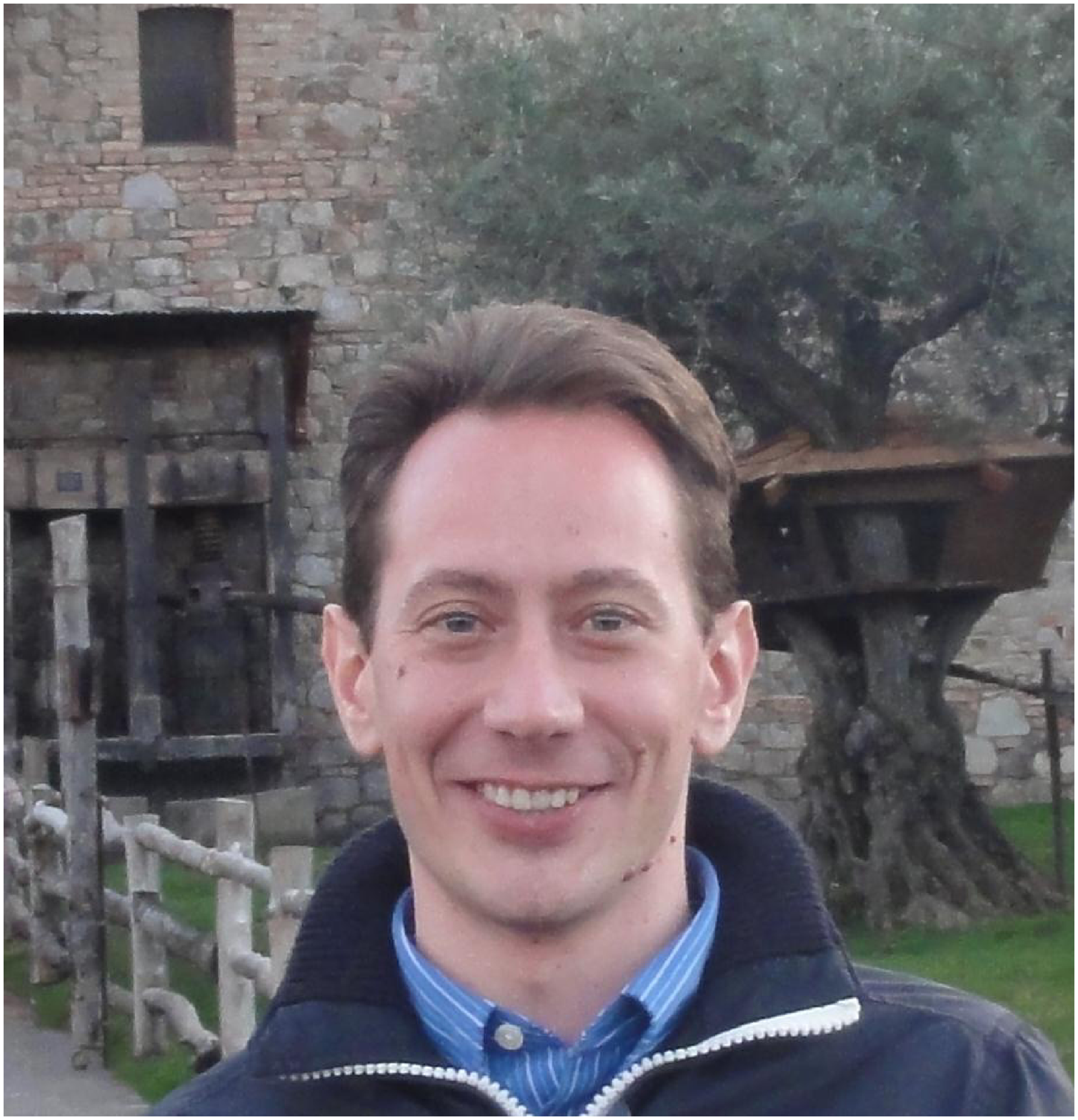}}]{Salvatore Talarico}
\fi
received the BSc and MEng degrees in electrical engineering from University of Pisa, Italy, in 2006 and 2007 respectively. From 2008 until 2010, he worked in the R$\&$D department of Screen Service Broadcasting Technologies (SSBT) as an RF System Engineer. He is currently a research assistant and a Ph.D. student in the Lane Department of Computer Science and Electrical Engineering at West Virginia University, Morgantown, WV. His research interests are in wireless communications, software defined radio and modeling, performance and optimization of ad-hoc and cellular networks.
\end{IEEEbiography}

\end{document}